\documentclass[journal=ancac3,manuscript=article]{achemso}

\usepackage[version=3]{mhchem} 
\usepackage[T1]{fontenc}       
\usepackage{float}
\usepackage{graphicx}
\usepackage{bm}
\usepackage{epsfig}
\usepackage{epstopdf}
\usepackage{amsmath}
\usepackage{amstext}
\usepackage{amssymb}

\author{Jayakanth Ravichandran}
\email{jayakanr@usc.edu}
\affiliation{Department of Physics, Harvard University, Cambridge, MA 02138, USA}
\altaffiliation{Current Address: Department of Chemical Engineering and Materials Science, University of Southern California, Los Angeles, CA 90089, USA}
\phone{+1 213-740-0453}
\author{Claudy Rayan Serrao}
\affiliation{Department of Materials Science and Engineering, University of California, Berkeley, CA 94720, USA}
\author{Dmitri Efetov}
\affiliation{Department of Physics, Columbia University, New York, NY 10027, USA}
\altaffiliation{Current Address: Department of Electrical Engineering, Massachusetts Institute of Technology, Cambridge, MA 02138, USA}
\author{Di Yi}
\affiliation{Department of Materials Science and Engineering, University of California, Berkeley, CA 94720, USA}
\author{Yoon Seok Oh}
\affiliation{Rutgers Center for Emergent Materials and Department of Physics and Astronomy, Piscataway, New Jersey 08854, USA}
\author{Sang-Wook Cheong}
\affiliation{Rutgers Center for Emergent Materials and Department of Physics and Astronomy, Piscataway, New Jersey 08854, USA}
\author{Ramamoorthy Ramesh}	
\affiliation{Department of Materials Science and Engineering, University of California, Berkeley, CA 94720, USA}
\alsoaffiliation{Lawrence Berkeley National Laboratory, Berkeley, CA 93720, USA}
\alsoaffiliation{Oak Ridge National Laboratory, Oak Ridge, TN 37831, USA} 
\author{Philip Kim}
\email{pkim@physics.harvard.edu}
\affiliation{Department of Physics, Harvard University, Cambridge, MA 02138, USA}
\altaffiliation{Current Address: Department of Physics, Harvard University, Cambridge, MA 02138, USA}
\phone{+1 617-496-0714}

\title{Ambipolar Transport and Magneto-resistance Crossover in a Mott Insulator, Sr$_{2}$IrO$_{4}$}


\begin{document}









\begin{abstract}
  Electric field effect (EFE) controlled magnetoelectric transport in thin films of undoped and La-doped Sr$_{2}$IrO$_{4}$ (SIO) were investigated under the action of ionic liquid gating. Despite large carrier density modulation, the temperature dependent resistance measurements exhibit insulating behavior in chemically and EFE doped samples with the band filling up to 10\%. The ambipolar transport across the Mott gap is demonstrated by EFE tuning of the activation energy. Further, we observe a crossover from a negative magnetoresistance (MR) at high temperatures to positive MR at low temperatures. The crossover temperature was around $\sim$80-90 K, irrespective of the filling. This temperature and magnetic field dependent crossover is qualitatively associated with a change in the conduction mechanism from Mott to Coulomb gap mediated variable range hopping (VRH). This explains the origin of robust insulating ground state of SIO in electrical transport studies and highlights the importance of disorder and Coulombic interaction on electrical properties of SIO. 
\end{abstract}


Electric field effect (EFE)~\cite{Ahn:2006ts} is a powerful tool to study the effect of band filling on transport properties, without creating doping induced disorder. In particular, recent progress in the use of ionic liquids for electrostatic modification of surfaces has extended the limits of EFE doping close to $\sim$10$^{15}$~cm$^{-2}$~\cite{Misra:2007dg,Ueno:2011cg,Leng:2011jw,Bollinger:2011eq,Efetov:2010fu}. The thin ($\sim$~1nm) electric double layer formed at the interface of the channel and the ionic liquid allows us to accumulate or deplete such a large carrier density. This method has enabled experimental probing of a wide range of the electron density dependent phase diagrams~\cite{Bollinger:2011eq,Leng:2011jw} and phases inaccessible by chemical doping~\cite{Ueno:2011cg}. Thus, EFE doping with ionic liquids provides a good experimental platform to explore the electronic phase diagram of a variety of material systems without significant chemically induced disorder.

Transition metal oxides (TMOs) are excellent model systems for Mott insulators, where localized $d$-orbitals cause strong electron correlation effects~\cite{Brandow:1977ip,Imada:1998er}. As we move down the periodic table, from 3-$d$ to 5-$d$, TMOs show weaker electron correlation but larger spin-orbit coupling, due to higher atomic numbers. Among the 5-$d$ TMOs, Ruddlesden-Popper series of iridates such as Sr$_{n+1}$Ir$_{n}$O$_{3n+1}$ ($n>0$), specifically Sr$_{2}$IrO$_4$ (SIO), show an unusually robust insulating ground state, which persists despite chemical doping and applied pressure~\cite{Klein:2008tf,Haskel:2012ix}. The nature and origin of the insulating nature of SIO remains controversial~\cite{Kim:2008tp,Moon:2009bx,Hsieh:2012ch,Dai:2014ey,Li:2013ti}. This compound also shows a weak ferromagnetic moment~\cite{Cao:1998js}, attributed to its canted antiferromagnetic ground state~\cite{Boseggia:2013jo,Kim:2012cr,Fujiyama:2012ig}. Due to the interplay between electron correlation, spin-orbit coupling, magnetism and crystal field effects, understanding the nature of insulating state of SIO remains complicated and yet interesting. Existing transport studies report insulating and/or non-Ohmic behavior in SIO~\cite{Klein:2008tf,Cao:1998js,Korneta:2010co} and thus have provided little insight into the role of disorder and transport mechanisms of the insulating state. Moreover, predictions of superconductivity in the electron doped SIO and its similarities to La$_2$CuO$_4$~\cite{Wang:2011cm} invites a thorough transport study  to identify a route to achieve metallicity in Mott insulating SIO.

\section{Results and discussion}

In this article, we study the evoluation of magneto-transport properties upon EFE doping of undoped and La-doped SIO thin films (5\% La - La$_{0.05}$Sr$_{1.95}$IrO$_{4}$ and 10\% La - La$_{0.1}$Sr$_{1.9}$IrO$_{4}$) at low temperatures. EFE doping was achieved using ionic liquid gating. We measured magnetotransport under ionic liquid gating for the chemically undoped samples, where an ambipolar transport characteristics is observed. We found that the low temperature transport is always dominated by VRH mechanism, accompanied by a MR crossover at $\sim$80-90~K. Both these observations were unaffected by the changes in band filling. 

\subsection{Temperature Dependent Resistance Measurements}
\begin{figure}[t]
\epsfbox{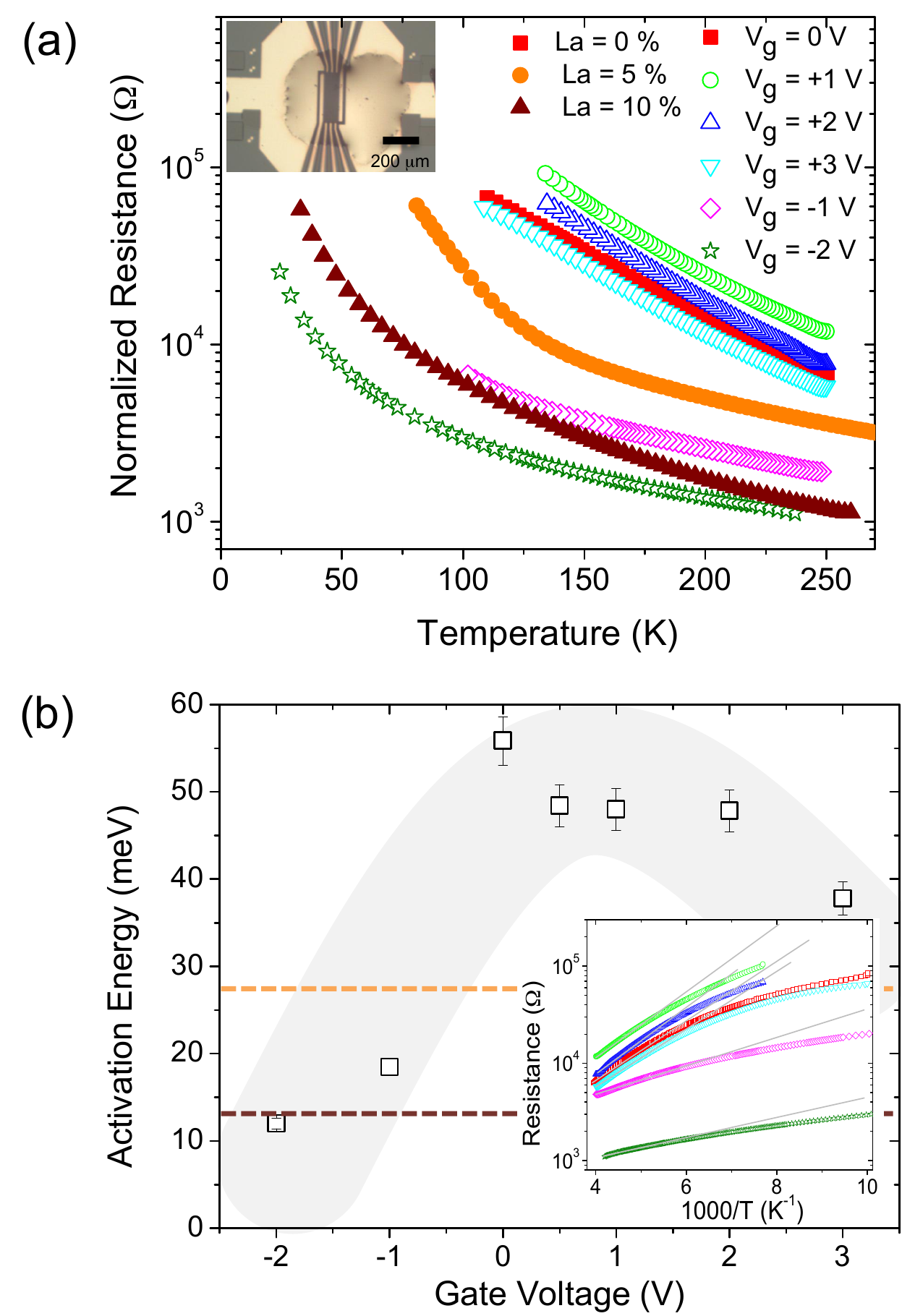}
\caption{\label{Fig:fig1} (a) Temperature dependent resistance of an undoped 10 nm thick SIO film with ionic liquid gating (for different applied gate voltages, $V_g$) and chemically doped films (5 and 10\% La). The resistance values of chemically doped films were normalized to a 10 nm thick film for comparison. The inset shows the optical image of the Hall bar device with the ionic liquid. (b) The extracted activation energy (black open squares) is shown as a function of $V_g$. The shaded curved band is used as a guide-to-the-eye. The inset shows the channel resistance plotted as a function of inverse of temperature. The symbols and the color codes used are the same as in (a). The grey lines show the fits for activated transport, whose slopes were used to obtain the activation energy at the high temperature regime. The orange and brown horizontal dotted lines corresponds to activation energy for 5\% and 10\% La doped samples, respectively.}
\end{figure}

We measured the resistance $R$ of the samples as a function of the temperature $T$ in the range between 1.5-200~K. We applied gate voltage ($V_g$) to the side gate in contact with ionic liquid at 260~K and subsequently the sample was cooled at a fixed $V_g$. Below 220~K, the ionic liquid freezes out completely and stable electrical measurements become possible. The gate modulation in the range of -2~V to 3~V was achieved without significant degradation of samples. $V_g$ applied outside of this regime caused electrochemical reactions, resulting in irreproducible $R(T)$. \ref{Fig:fig1} (a) shows $R(T)$ for SIO films doped chemically with La (filled symbols) and undoped samples using an ionic liquid gate (open symbols). All samples exhibit $dR/dT<0$, i.e., an insulating behavior, regardless of chemical doping (up to 10\% La doping). 

\setlength{\epsfxsize}{0.7\columnwidth}

The resistance of ionic liquid gated sample changes rapidly with decreasing $T$, whose behavior can be modulated by $V_g$. To quantify this gate dependence, we replot $R(T)$ in an Arrhenius form for a fixed $V_g$ (inset of \ref{Fig:fig1}~(b)). The temperature dependence of these samples can be divided broadly into two regimes: (i) high temperature activated regime ($T\gtrsim150$~K), where $R(T)$ follows an activated (Arrhenius) behavior;  and (ii) low temperature VRH regime ($T\lesssim150$~K), where $R(T)$ deviates from the Arrhenius behavior and exhibits hopping conduction. The activation energy, $E_a$, can be obtained from the slope of the linear fit in the high temperature regime of the Arrhenius plot. \ref{Fig:fig1}~(b) shows the plot of $E_a$ versus $V_g$. Interestingly, $E_a$ drops as $|V_g|$ increases with a peak value of $\sim 50$~meV at $V_g=0$~V. This ambipolar behavior suggests that the Fermi level $E_F$ of the undoped ($V_g=0$~V) sample is close to the middle of the gap. Hall measurement (data not shown) also indicates that $V_g>+1 V$ ($< +1 V$) corresponds to n-type (p-type) carriers and the undoped samples ($V_g=0$) are weakly p-type. The weak p-type nature of the undoped sample also results in the asymmetry of the activation energy as a function of $V_g$. Thus, the bell-shaped plot $E_a(V_g)$ establishes ambipolar transport behavior in the Mott insulating SIO.  Previously, demonstration of ambipolar doping in a Mott insulator was realized only by chemical doping across many samples\cite{Segawa:2010gu}. In our experiment, similar ambipolar doping of a Mott insulator (SIO) is demonstrated in a single sample \textit{via} electrochemical modulation. 

\setlength{\epsfxsize}{0.8\columnwidth}
\begin{figure}[t]
\epsfbox{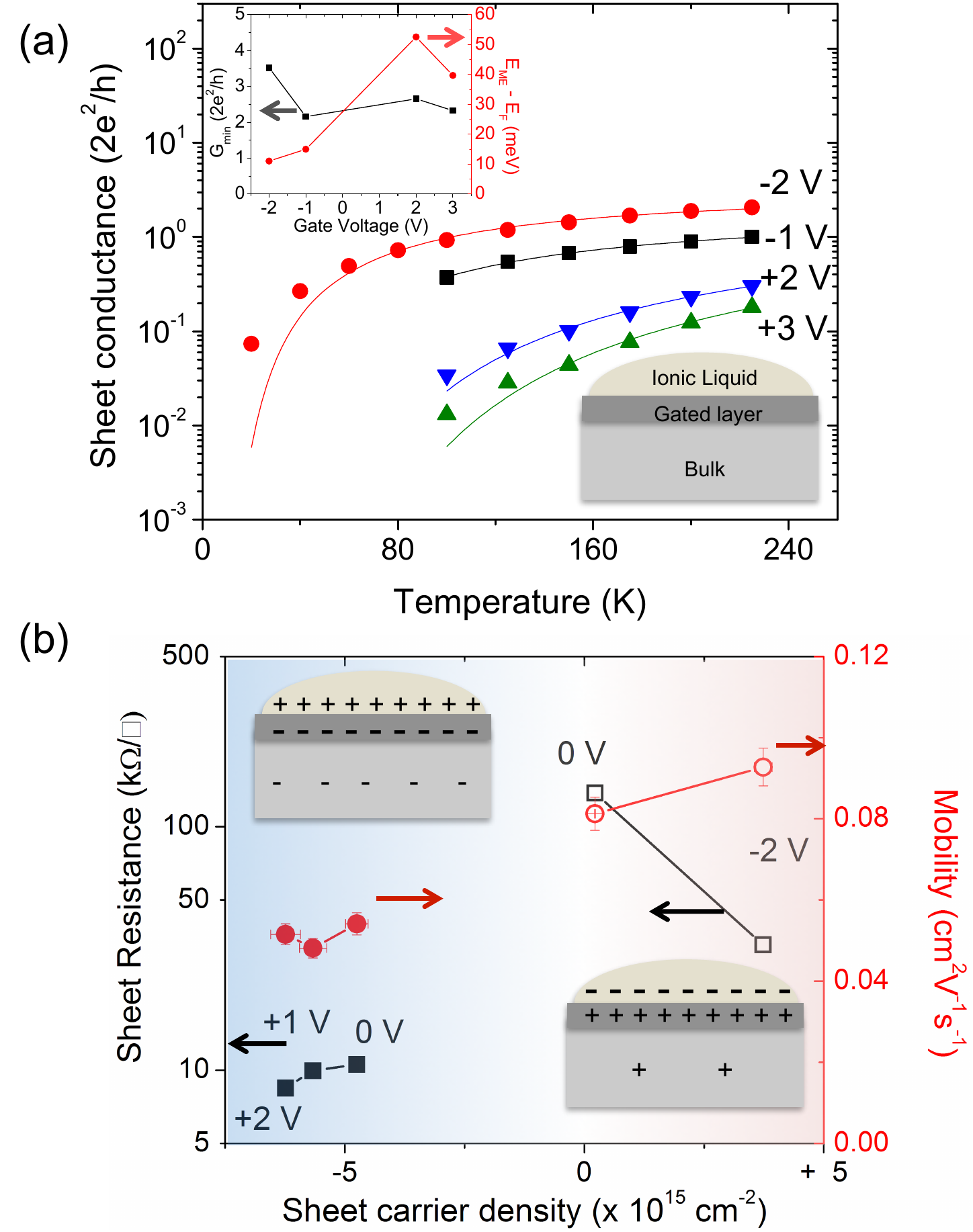}
\caption{\label{Fig:fig2} (a) Temperature dependent sheet conductances of the ionic liquid gated surface layers measured at different gate voltages. The conductance of the surface layer $\Delta G$ was obtained by subtracting the conductance of the ungated portion of the samples (see the lower inset for a schematic bi-layer model). The solid lines are fits using activation to the mobility edge (see text).  The upper inset shows the variation of the $G_{min}$ (black squares)  and $E_{ME}-E_{F}$ (red circles) with the applied gate voltage. (b) Gate modulated resistivity (black squares) and Hall mobility (red circles) for a 10 nm thick 10\% La doped SIO thin film (filled symbols) and a 25 nm thick undoped SIO thin film (open symbols) as a function of the Hall sheet carrier density (modulated by the gate voltage). The data were obtained at similar temperatures of 190 and 150~K respectively. The inset shows the schematic diagram of free and bound charge distributions in the samples. 
}
\end{figure}

\subsection{Effect of ionic liquid gating on film conductance}

In order to understand the nature of the gated surface layer, we used a simple two layer model as depicted in the lower inset of \ref{Fig:fig2} (a). Hall measurements show that the undoped bulk sample has a weak p-type doping (\ref{Fig:fig2} (b)), with a carrier density $n \approx 2.5\times 10^{14}$~cm$^{-3}$. Upon electrolyte gating, EFE induced charges accumulate on the surface layer, which has a different conductivity as compared to the bulk of the film. For the hole accumulation side ($V_g<0$), the conductance of the surface layer, $G_s$, can be estimated from the difference of the conductance with reference to $R_0=R(V_g=0\mbox{V})$: $G_s(V_g)=R(V_g)^{-1}- R_0^{-1}$.  

 In the case of electron doping ($Vg>0$), a depletion layer is formed first on the surface and then gradually becomes electron doped (n-type) for higher $V_g$. Therefore $R(V_g)$ increases initially when $V_g$ increases as shown in \ref{Fig:fig1}(a) and then starts decreasing for higher $V_g$. For undoped samples, $R(V_g)$ is the largest for $V_g\approx +1$~V. The thickness of depletion layer, $\ell_{d}$, thus can be estimated from $\ell_d=(R(Vg=1~\mbox{V})/R_0-1)d \approx 4$~nm, where $d=10$~nm is the thickness of the sample. Using $\ell_d$, we can estimate $G_s$ for this inversion regime by subtracting the sheet conductance of the depleted layer using~: $G_s= R(V_g)^{-1}- \left(\frac{ d-\ell_d }{d}\right)R_0^{-1}$. \ref{Fig:fig2}~(a) shows $G_s$ ($T$) for fixed $V_g$. The conductance steeply drops as $T$ decreases, which is expected from the transport through localized states. We employ a simple activation model to mobility edge to describe the temperature dependent transport through the surface conduction layer:  $G_s = G_{min}\exp\left[\frac{-\left(E_{ME}-E_{F}\right)}{k_{B} T}\right]$,
where $E_{ME}$, $E_{F}$, and $G_{min}$ are the mobility edge, Fermi energy, and Mott minimum conductance, respectively. Typically such an analysis is performed on semiconductors which exhibit disorder induced mobility edge~\cite{Arnold:1974ke}, where characteristic carrier density and temperature dependence of the mobility is observed. Using this simple activation formula, we fit the experimental data of temperature dependent $G_s$ for fixed $V_g$ (solid lines in \ref{Fig:fig2}~(a)) and obtain $E_{ME}$ - $E_{F}$ and $G_{min}$ for different gate voltages, which are shown in the upper inset of \ref{Fig:fig2}~(a). The activation fits are reasonable for most of gate voltages at higher temperature ranges ($T\gtrsim$~150~K),  but starts deviating at lower temperatures, likely indicating the limitations of such a model at low temperatures. The obtained values of $\Delta E=E_{ME}$ - $E_{F}$ are similar to the activation energy, $E_a$, and the $G_{min}$ is $\sim$ few $e^2/h$, as it is expected for 2D localization~\cite{Tsui:1974js}.
While the disorder driven localization seemingly explains the observed behavior of $G_s$, there are a few factors that are not captured by a simple localization theory. \ref{Fig:fig2}~(b) shows that the measured Hall mobility remains roughly constant ($\sim$0.08~cm$^2$/Vsec), despite that $G_s$ and $\Delta E$ decrease almost an order of magnitude for a wide range of carrier density (both electrons and holes). This filling independent mobility is not expected for disorder induced localization, where one expects a steep increase of mobility as $E_F$ is getting close to the mobility edge~\cite{Arnold:1974ke}. We also make a note on the strain effect on both the insulating state and the EFE on the transport properties. Although we have measured the transport properties of 25 nm thick SIO films on various substrates such as STO, LSAT, DSO (DyScO$_{3}$), and GSO (GdScO$_{3}$), which exert different amount of strain on SIO, the insulating nature of the material persists in all samples. Despite the qualitative similarity, the films showed subtle changes in the resistivity. There are a few reports on the effect of strain on the optical and transport gap of SIO in the literature~\cite{Nichols:2013fr,RayanSerrao:2013em}. The detailed study of strain effect, however, is beyond the scope of the current investigation.

\begin{figure}[t]
\epsfbox{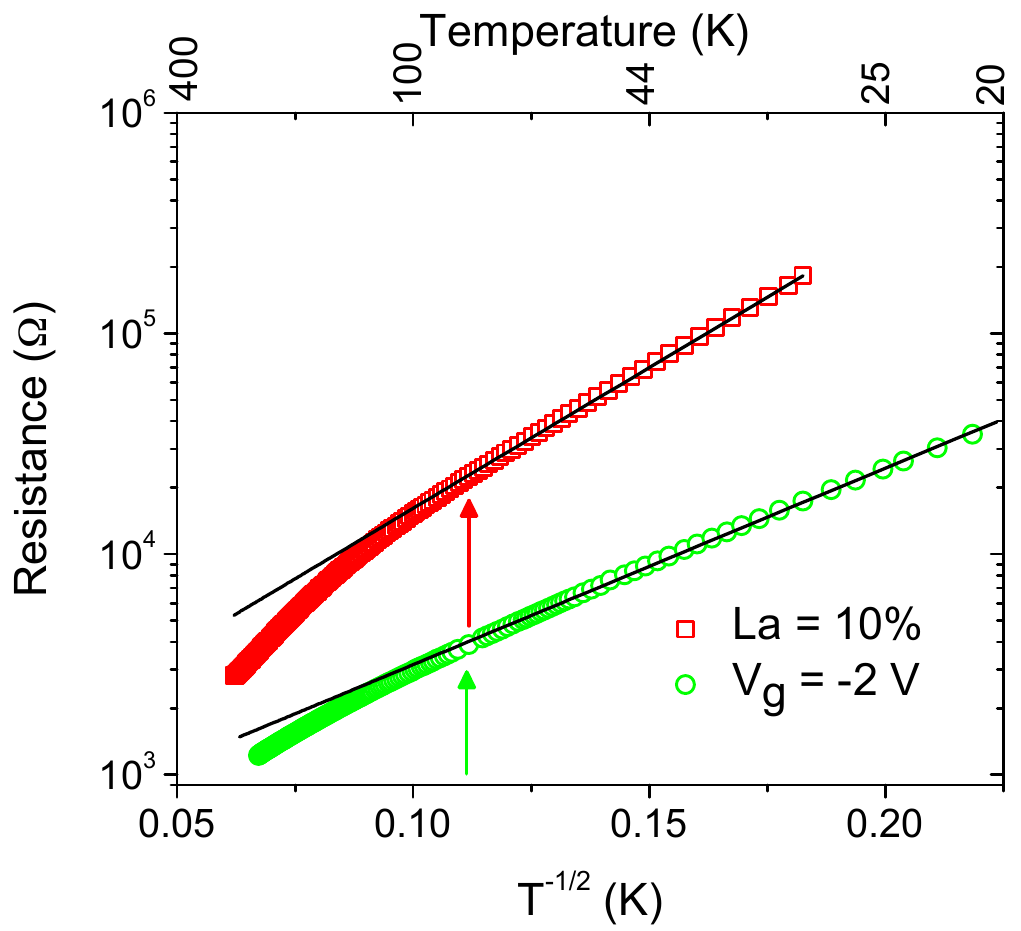}
\caption{\label{Fig:fig6}The temperature dependence of resistance for the two samples used for electron doped La$_{0.1}$Sr$_{1.9}$IrO$_{4}$ thin film and ionic liquid gated SIO (hole doped with $V_{g} = -2$~V) are shown as red squares and green circles respectively. The Efros-Shklovskii (ES) VRH fit for the data is shown as black lines. The vertical arrows mark where the fit shows a deviation. The resistance values for the 10\% La doped sample (red squares) are scaled up by a factor of 5 to avoid overlap between the plots.
}
\end{figure}

\subsection{Temperature Dependence of Resistance in the Variable Hopping Regime}

The low temperature hopping regime can be dominated by either defect mediated hopping, known as Mott VRH~\cite{ Mott:1969il}  or by Coulombic interactions, known as Efros-Shklovskii (ES) VRH~\cite{ Efros:1975he}. Temperature dependent VRH resistivity can be described by $\rho$ = $\rho_{0}\exp{(T_{0}/T)^{1/\alpha}}$, where $T_{0}$ is a characteristic hopping energy scale and $\alpha$ is the exponent, dependent on dimension and type of VRH~\cite{Mott:1969il,Efros:1975he}, where $\alpha = 4$ for Mott VRH and $\alpha=2$ for ES VRH. \ref{Fig:fig6} shows the resistance of electron doped SIO with 10\% La (red squares) and hole doped SIO with $V_{g}$ = -2 V (green circles) plotted against inverse square root of temperature. The data shows excellent agreement at low temperatures with $T_{0}$ as a fitting parameter. We obtain $T_{0}$ values of 860~K and 420~K for the electron and hole doped cases respectively. At higher temperature, the ES-VRH deviates from the measured data around $\sim$ 80~K as marked by the vertical arrows in the graph. The $T_{0}$ values estimated above are comparable to the values reported elsewhere~\cite{Lu:2014kj}. These values are also in agreement with the temperature criterion necessary to observe ES-VRH presented in~\cite{Rosenbaum:1991hx}, where we obtain $T<T_{0}/24.6$~$\sim$ 20-50~K, within the ES-VRH regime in our experiments. Also, we note that the crossover temperature ($T_{cr}$$\sim$ 80~K) is much higher than typical $T_{cr}$ in disordered semiconductors~\cite{Agrinskaya:1995wr,Rosenbaum:2001kb,Agrinskaya:1997fp}, but similar Mott-ES VRH crossover and $T_{cr}$ has been reported in indium oxide~\cite{Lu:2014kj,Rosenbaum:1991hx} and most recently, undoped SIO~\cite{Lu:2014kj}. 

\begin{figure*}[t]
\epsfbox{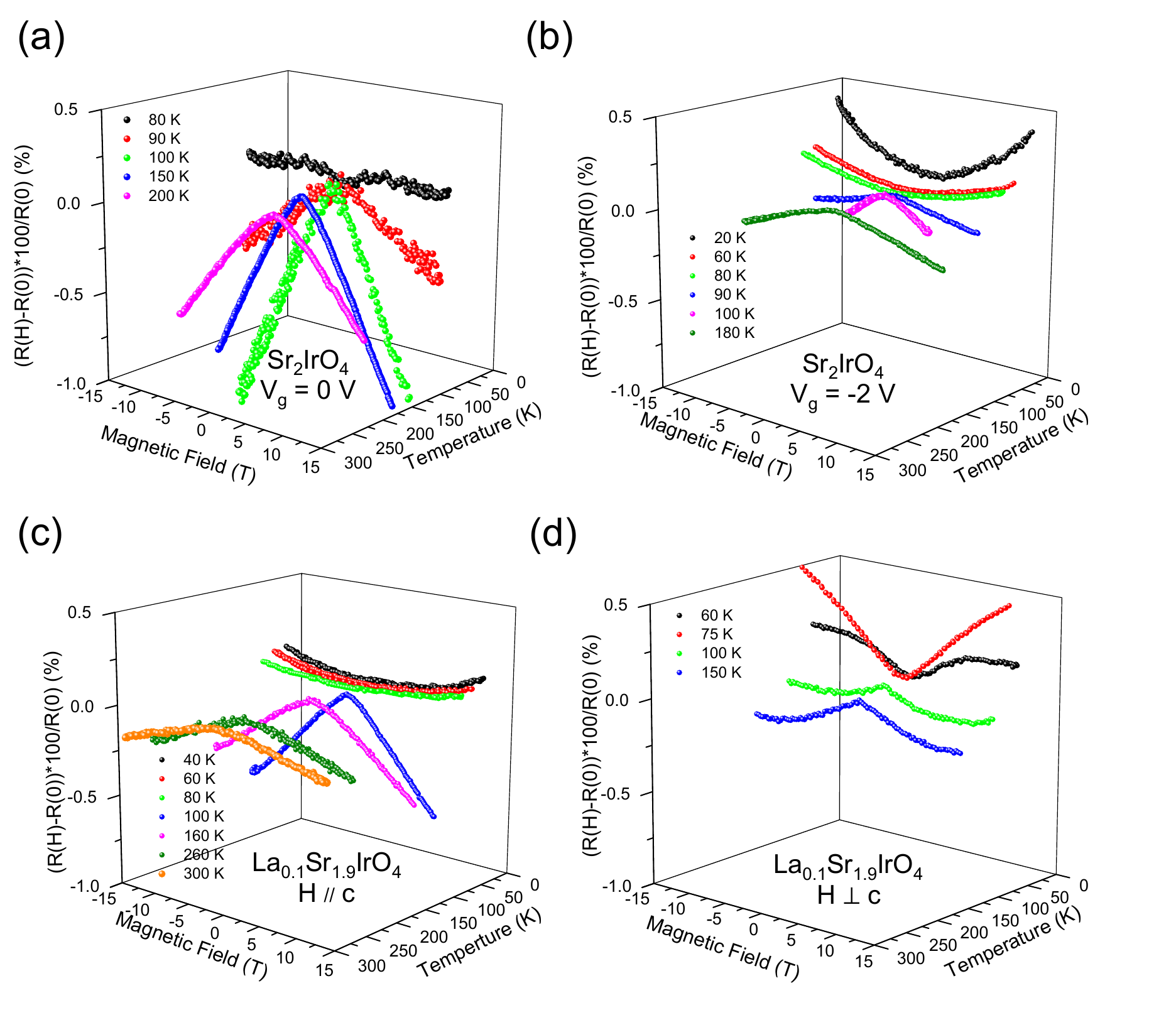}
\caption{\label{Fig:fig3} The measured MR as a function of both applied magnetic field and temperature for (a) undoped SIO, (b) ionic liquid gated SIO (hole doped with $V_{g} = -2$~V), (c) electron doped La$_{0.1}$Sr$_{1.9}$IrO$_{4}$ thin film, and (d) electron doped La$_{0.1}$Sr$_{1.9}$IrO$_{4}$ thin film with applied field parallel to $a-b$ plane. For all the other three cases, except (d), the magnetic field was applied parallel to $c$-axis.}
\end{figure*}

\subsection{Magnetoresistance Crossover in the Hopping Regime}

 As the temperature range used in the above temperature dependence analysis is rather limited (at best an order of magnitude change in temperature), any conclusion of the nature of hopping conduction should be supported by a stronger evidence. Hence, In order to clarify the nature of the hopping conduction in SIO, we performed temperature dependent MR measurements. MR has been studied extensively in the hopping conduction regime for several semiconductors and alloys~\cite{Agrinskaya:1995wr,Rosenbaum:2001kb,Zhang:1992et,Srinivas:2001jc,Rentzsch:1997ic}, where the temperature and field dependence of MR has been used to explain the low temperature transport mechanism. \ref{Fig:fig3} shows the transverse MR for (a) undoped, (b) ionic liquid gated hole doped, and (c) chemically electron doped samples, and (d) longitudinal MR for electron doped samples. Both electron and hole doped samples demonstrate an abrupt MR sign change at $\sim 80-90$~K. The aforementioned Mott-ES crossover was also observed in a similar temperature range. At low temperatures, the MR exhibits a positive, quadratic scaling with the magnetic field, but at high temperatures, the MR shows a negative, linear scaling with the magnetic field. An abrupt change in sign and field dependence of MR at such high temperatures is unusual, which deserves careful analysis. Before we discuss the details of this crossover, we first focus on understand the limiting MR behavior. The high temperature linear negative MR in the VRH regime is often attributed to the quantum interference between the hops~\cite{Nguen:1985uv}, which was experimentally observed in disordered semiconductors and quasicrystalline alloys~\cite{Zhang:1992et,Srinivas:2001jc}. The low temperature positive MR can be attributed to several possibilities including (i) the shrinking of electronic impurity wave functions under a magnetic field~\cite{Nguen:1985uv}; (ii) spin dependent MR (Kamimura effect)~\cite{Kamimura:1983gd} or (iii) superconducting or magnetic fluctuations~\cite{Maki:1968es}. We eliminate the role of superconducting fluctuations (i.e., (iii) above) as we are far from the metallic limit. We also exclude the possibility of a magnetic fluctuation based mechanism ((ii) above), since a decreasing MR with temperature is expected as we approach the ordering temperature. Moreover, we performed magnetization measurements on the undoped and 10\% La doped samples, which exhibit a magnetic  transition at temperatures of $\sim 240$~K and $\sim 150$~K respectively (Supporting Information, Figure S1 and Section II) . This suggests that the magnetic transition temperature does not correlate with the observed MR sign $T_{cr}$. Thus, we attribute the shrinking of the electronic impurity wave function as a probable cause for the observed positive MR at low temperatures and also provide supporting arguments, based on detailed MR measurements about the $T_{cr}$.  

\setlength{\epsfxsize}{1\columnwidth}
\begin{figure}[t]
\epsfbox{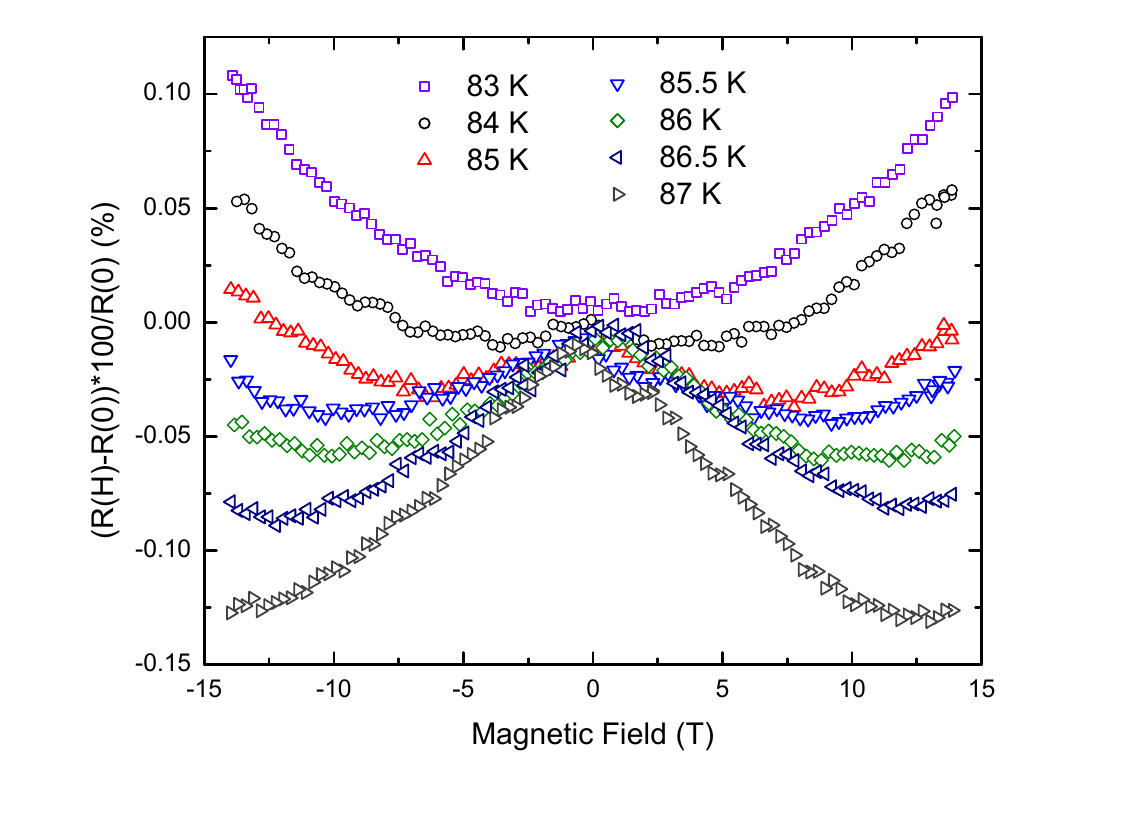}
\caption{\label{Fig:fig5} The MR measured as a function of the magnetic field at different temperatures (83 - 87 K) right around the MR sign change for the electron doped La$_{0.1}$Sr$_{1.9}$IrO$_{4}$ thin film (shown in \ref{Fig:fig3}~(a)). The graph clearly highlights a MR crossover from negative to positive values, which is sensitively dependent on both the magnetic field and temperature. 
}
\end{figure}

\setlength{\epsfxsize}{0.7\columnwidth}
\begin{figure}[t]
\epsfbox{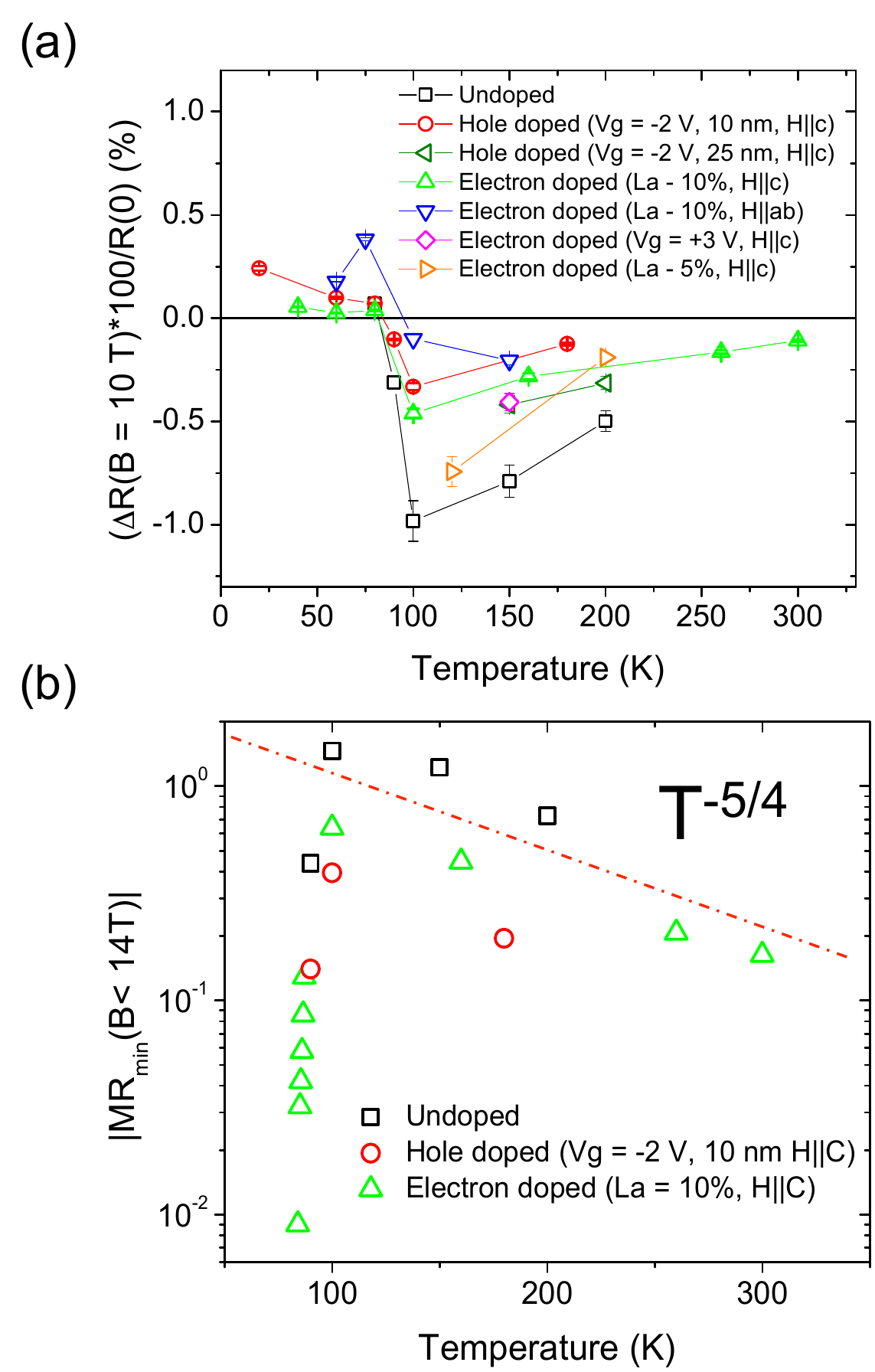}
\caption{\label{Fig:fig4}~(a) The MR measured at the magnetic field of 10~T as a function of temperature for different n-type and p-type doping concentrations. The graph clearly highlights a MR crossover from negative to positive values at about $\sim 80-90$~K. The graph also shows high temperature negative MR for other doping (La 5\%) and ionic liquid gated samples with different gate voltages used in this study. (b) $MR_{min}$ as a function of temperature for undoped (black squares), hole doped (open circles) and electron doped (open triangles) samples. The MR in the high temperature regime exhibits a T$^{-5/4}$ scaling, where as in the low temperature regime, MR shows a fast decay. The red dotted lines are used a guide-to-the-eye for the high temperature power law scaling.}
\end{figure}

Further supporting evidence of Mott-ES VRH crossover can be provided by detailed analysis of the crossover behavior of MR. \ref{Fig:fig5} clearly shows that the field and temperature dependent MR crossover is observed in 10\% La doped samples over a narrow range of temperatures from 83 - 87 K. This behavior is strikingly similar to the MR crossover reported in the hopping regime of several disordered semiconductors and alloys~\cite{Agrinskaya:1995wr,Rosenbaum:2001kb,Zhang:1992et,Srinivas:2001jc,Rentzsch:1997ic}, although only qualitatively. \ref{Fig:fig4}~(a) shows the variation of MR at an applied field of 10~T, highlighting the experimental observation that the sign change is qualitatively similar for a range and type of band filling (both n-type and p-type). Although the magnitudes of MR are quite different, interestingly, the MR sign change appears in all of our samples with nearly similar crossover temperature $T_{cr}\sim~$80~K. Similar temperature dependent MR behavior, which extended beyond the crossover region, was reported in highly compensated n-GaAs~\cite{Rentzsch:1997ic}, further suggesting the qualitative similarities of MR crossover in these systems.

The MR crossover behavior in disorder systems can often be associated with a Coulomb gap opening~\cite{Agrinskaya:2010fq,Agrinskaya:1995wr}. \ref{Fig:fig4}~(b) shows the extracted minimum MR (for a field of $0<H<14$~T), $MR_{min}=\left[ R(B)/R(0)-1\right]_{min}$, as a function of temperature. At high temperatures, the absolute value of $MR_{min}$ increases as $T$ decreases, but then, for $T<T_{cr}\sim$80~K, $MR_{min}$ drastically decreases with decreasing $T$. Considering both the interference of different tunneling paths with intermediate scattering centers and the deformation of the wave functions of the VRH sites, $MR_{min}(T)$ is a sensitive probe to the characteristic energy of the VRH (W), Coulomb gap energy ($W_C$) and the potential depth of the ionized impurities ($\delta$)~\cite{Agrinskaya:2010fq,Agrinskaya:1995wr}. Across the ES-Mott $T_{cr}$, $MR_{min}(T)$ exhibits the power law dependence on temperature with different exponents. As it is indicated by a dotted line in \ref{Fig:fig4}~(b), for $T>T_{cr}$, $MR_{min}$ follows closely $\sim T^{-5/4}$ scaling independent of band filling, corresponding to $W_C$, $\delta$ $<$ W, suggesting a Mott VRH dominated transport in this temperature range~\cite{Agrinskaya:1995wr}. For $T<T_{cr}$,  $MR_{min}$ decreases with decreasing T, signifying the suppression of negative MR, which is characteristic of the ES-VRH regime (Supporting Information, Section I). Thus, this MR crossover is attributed to the transition from Mott VRH to ES-VRH, due to the opening of Coulomb gap. We note that the unusually fast suppression of the negative MR in the low temperature regime exhibits a power law decay much faster than the expected theoretical models~\cite{Agrinskaya:1995wr,Agrinskaya:2010fq}. Such a discrepancy, however, if often observed in disordered semiconductors or alloys~\cite{Agrinskaya:1995wr,Rosenbaum:2001kb,Agrinskaya:1997fp}. Based on both phenomenological and numerical calculations presented in~\cite{Agrinskaya:1997fp}, we rule out the possibility of the dominant role of spin scattering or spin-orbit coupled scattering in causing suppression of negative MR. Theoretically, the formation of Coulomb gap during ES-VRH is the consequence of electron interaction and is also shown to be independent of the band filling~\cite{Davies:1986ct}, in agreement with our experimental observations, explaining the universal crossover temperature across all doping in this system.Detailed analysis of the $MR_{min}$ crossover is given in the supplementary online information (Section I).

\section{Conclusion}

In conclusion, we have demonstrated that ionic liquid gating can be used to realize ambipolar transport in thin films of SIO. Despite the large carrier injection, the insulating state persists for both electron and hole doping attainable in our experiments. We observe a robust MR sign change, signifying a crossover from Mott VRH to ES-VRH. The origin of the persistent transport insulating state in SIO at the low temperature regime is attributed to the opening of Coulomb gap.

\section{Methods}

~~~~~\textbf{Thin film synthesis:} Thin films of SIO (10--25 nm thick) were grown epitaxially on (LaAlO$_3$)$_{0.3}$-(Sr$_2$AlTaO$_6$)$_{0.7}$ (LSAT) and SrTiO$_3$ (STO) substrates using pulsed laser deposition, as discussed elsewhere~\cite{RayanSerrao:2013em}. The films were grown at 850$^{\circ}$C inan oxygen partial pressure of 1 mTorr. A KrF (248 nm) excimer laser with a fluence of 1.1 J/cm$^{2}$ was used to ablate a stoichiometric ceramic target of Sr$_{2}$IrO$_{4}$ and after the growth the films were cooled down in 1 atm of oxygen partial pressure. The films were characterized using x-ray diffraction and transmission electron microscope for structural quality.  

\textbf{Device Fabrication:}The devices for transport measurements were fabricated in both Hall bar or van der Pauw device geometry. Electrodes for the samples and the side gate for electrolyte were fabricated using electron beam lithography with poly(methyl methacrylate) (PMMA) as the resist. Pd (10~nm)/Au (70~nm) contacts and side gate electrodes were deposited using electron beam metal evaporation. The films were then etched into Hall bar geometry, whose lateral size is $\sim$100~$\mu$m, using Ar ion milling. A small drop of ionic liquid, N, N-diethyl-N-(2-methoxyethyl)-N-methylammonium bis (trifluoromethyl sulphonyl)-imide (DEME TFSI), was applied on to the device with the liquid covering the entire channel and the side gates working as a counter electrode in an Argon filled glove box and quickly transferred to cryostat to avoid incorporation of moisture in the ionic liquid.

\textbf{Transport Measurements:} The transport measurements were carried out in a Quantum Design PPMS with a customized probe for low frequency ac measurements. The ac measurements were carried out using SRS 830 lock-in amplifier. The temperature sweep rate in the cryostat was typically set at 1 K/min. The resistance measurements were carried out between 1.5-200~K. The gate voltage was applied at 260~K and subsequently cooled to 220~K. Below 220~K, the ionic liquid freezes out completely and stable electrical measurements become possible. The applied gate voltage was restricted to -2~V to 3~V to avoid irreversible Faradaic processes in the device.  


\textit{Conflict of Interests:} The authors declare no conflicting financial interest. 


\textit{Acknowledgement.} The work at Harvard was supported by a DMREF grant from National Science Foundation (NSF)(DMR-1435487). The work at Columbia was supported a NSF MIRT
(DMR-1122594). The work at Berkeley was supported by the National Science Foundation through the Penn State MRSEC (D. Yi) and by the SRC FENA-FAME program through UCLA (C. R. Serrao). The work at Rutgers University was funded by the Gordon and Betty Moore Foundation’s EPiQS Initiative through Grant GBMF4413 to the Rutgers Center for Emergent Materials. We thank J.~T.~Heron for the help with magnetization measurements. The authors gratefully acknowledge several useful discussions with A.~Millis, I.~Aleiner, M.~Mueller, C.~Marianetti, J.~Liu and X.~Marti. 


\begin{suppinfo}

Derivation of $MR_{min}$ to establish the crossover and scaling relationships (Section I). Magnetization measurements on undoped and doped SIO to show the difference in the magnetic transition temperature and MR crossover temperature (Section II and Figure S1).

\end{suppinfo}


\providecommand*{\mcitethebibliography}{\thebibliography}
\csname @ifundefined\endcsname{endmcitethebibliography}
{\let\endmcitethebibliography\endthebibliography}{}

\end{document}